\documentclass[oneside,a4paper,10pt]{llncs}
\usepackage[english]{babel}
\usepackage{amssymb}
\usepackage{graphicx}
\usepackage{units}
\usepackage[usenames,dvipsnames]{xcolor}
\usepackage{fixltx2e}
\usepackage{booktabs}
\usepackage{setspace}
\usepackage{listings}
\usepackage{footmisc}
\usepackage{wrapfig}
\usepackage{paralist}
\usepackage{url}

\usepackage[a4paper,asymmetric,top=1.22in,bottom=1.6in,left=1.67in,right=1.39in]{geometry}

\let\subparagraph\paragraph
\usepackage{titlesec}
\usepackage{chngcntr}
\usepackage{hyperref}
\usepackage{tikz}
\usetikzlibrary{shapes,arrows,shadows,fit}

\titlespacing*{\section}{0pt}{1ex}{.5ex}
\titlespacing*{\subsection}{0pt}{1ex}{.5ex}
\titlespacing*{\subsubsection}{0pt}{0.5ex}{0ex}
\definecolor{shadecolor}{gray}{.95}                                                           

\urldef{\mailsa}\path|{becker, mneumair, asoehn}@rcs.ei.tum.de, samarjit@tum.de|
\newcommand{\keywords}[1]{\par\addvspace\baselineskip
\noindent\keywordname\enspace\ignorespaces#1}

\tikzstyle{every picture}+=[remember picture]

\begin{document}
\counterwithout{lstlisting}{chapter}

\mainmatter  

\title{\vspace*{-1cm}Approaches for Software Verification of an\\
  Emergency Recovery System for Micro Air Vehicles\thanks{The final publication will be available at Springer via \url{http://dx.doi.org/10.1007/978-3-319-24249-1_32}, in \emph{Computer Safety, Reliability and Security, 34th International Conference SAFECOMP 2015}, F.~Koornneef and C.v.~Gulijk (Eds.), Delft, Netherlands, 2015.}}

\titlerunning{SAFECOMP'15: Emergency Recovery System for MAVs}

%
%
\author{Martin Becker \and Markus Neumair \and Alexander S\"ohn \and Samarjit Chakraborty}
%

\institute{Institute for Real-Time Computer Systems\\Technische Universit\"at M\"unchen, Arcisstr. 21, 80333 Munich, Germany
\mailsa
}

%
%

\toctitle{SAFECOMP'15: Emergency Recovery System for MAVs}
\tocauthor{M. Becker et. al.}
\maketitle

\begin{abstract}\vspace{-1em}
  This paper describes the development and verification of a
  competitive parachute system for Micro Air Vehicles, in particular
  focusing on verification of the embedded software. We first
  introduce the overall solution including a system level failure
  analysis, and then show how we minimized the influence of faulty
  software. This paper demonstrates that with careful abstraction and
  little overapproximation, the entire code running on a
  microprocessor can be verified using \emph{bounded model checking},
  and that this is a useful approach for resource-constrained embedded
  systems.
  The resulting Emergency Recovery System is to our best knowledge the
  first of its kind that passed formal verification, and furthermore
  is superior to all other existing solutions (including commercially
  available ones) from an operational point of view.
\vspace{-.5em}
\end{abstract}

\section{Introduction}
In the recent years, Micro Air Vehicles (MAVs) such as quadrocopters, hexacopters, etc., are a rapidly growing class of airspace users. As of January 2015, we estimate the number of light MAVs ($<$\unit[5]{kg})
to be at least \unit[1.6]{million} in \emph{Europe}\footnote{Based on the
  number of DJI sales and their growing business figures over the last
  years. This is also supported by the number of and growth rate of
  registered MAVs at the federal agencies around Europe, and their
  estimated number of unreported vehicles \cite{SteerDaviesGleave2014}.}, possibly even one
magnitude higher due to the plethora of manufacturers and custom
builds. In comparison, this is more than
quadruple the number of aircraft in general aviation \emph{worldwide} \cite{Association2014}, and soon, if not already, the daily flying hours will also catch up, thanks to a growing number of civil use cases.

However, in contrast to aircraft in general aviation, MAVs are usually
not subject to in-depth safety considerations, but tend to have a high
probability of failure. This comes from the nature of these systems:
They are open for modifications, little analyzed, and often not fully
understood by their operator. Together with the omnipresence of those
vehicles, this results in a considerable potential of MAVs endangering
their environment.

\begin{figure}
  \centering
  \hspace*{2.5cm}\vspace*{-3mm}
  \begin{tikzpicture}[]
    \node [inner sep=0pt,above right] 
    {\includegraphics[height=4.5cm]{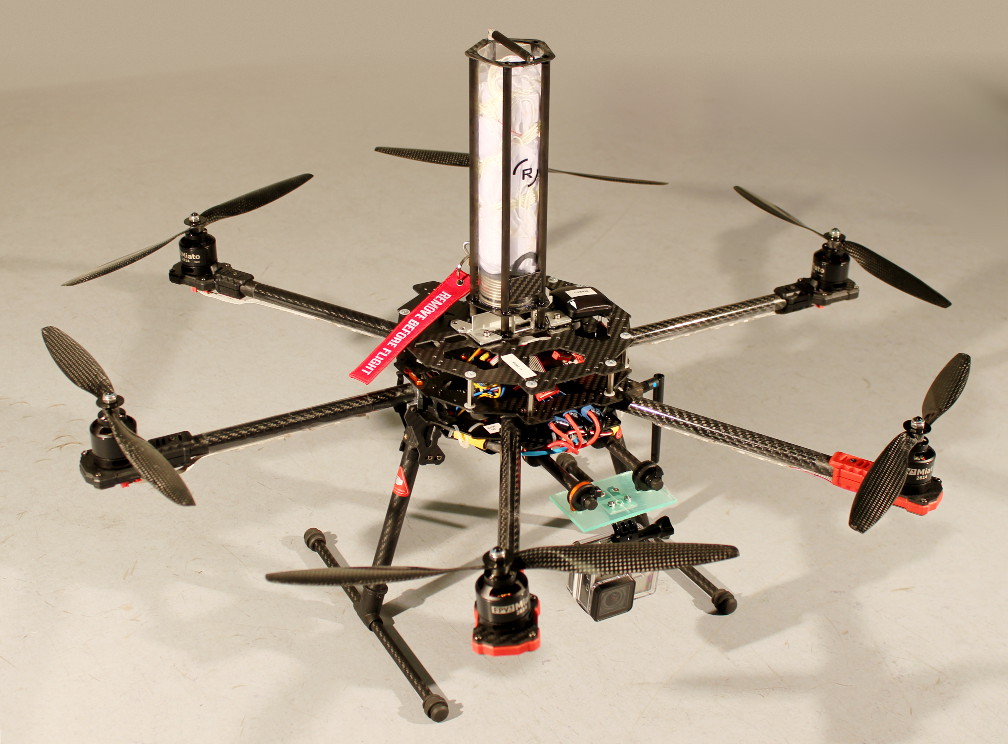}~\includegraphics[height=4.5cm]{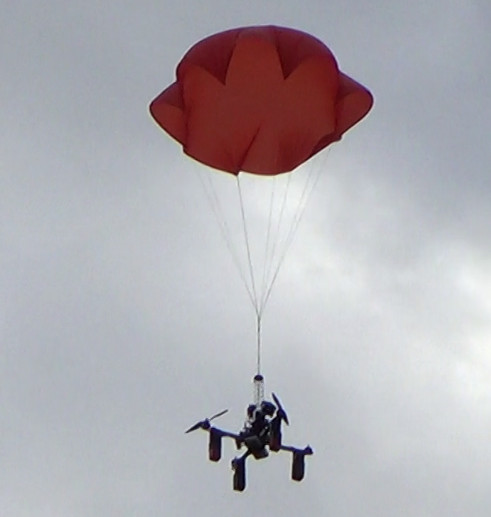}};
    \path (3,3.8) coordinate (tube)
    (2.3,2.3) coordinate (pin)
    (3,4.3) coordinate (sensor)
    (3,2.76) coordinate (spring);
  \end{tikzpicture}
\begin{tikzpicture}[overlay,>=latex, myarrows/.style={<-,MidnightBlue,thick}, ,mynodes/.style={pos=1.0,left,font=\footnotesize}]
        \path[myarrows] (tube) edge [] node [mynodes] (sd) {\footnotesize Parachute}  ++(-3.2,0);
        \path[myarrows] (sensor) edge [] node [mynodes] (sd) {\footnotesize Ejection Sensor}  ++(-3.2,0);
        \path[myarrows] (spring) edge [bend right=10] node [mynodes] (sd) {Spring}  ++(-3.18,.5);
        \path[myarrows] (pin) edge [bend left=20] node [mynodes] (sd) {\footnotesize Lock Pin}  ++(-2.5,-1.6);

\end{tikzpicture}
  \caption{Prototypes of our \emph{Emergency Recovery System} mounted on a hexacopter (left) and deployed on a quadrocopter (right).\label{fig:deploy}}
 \end{figure}
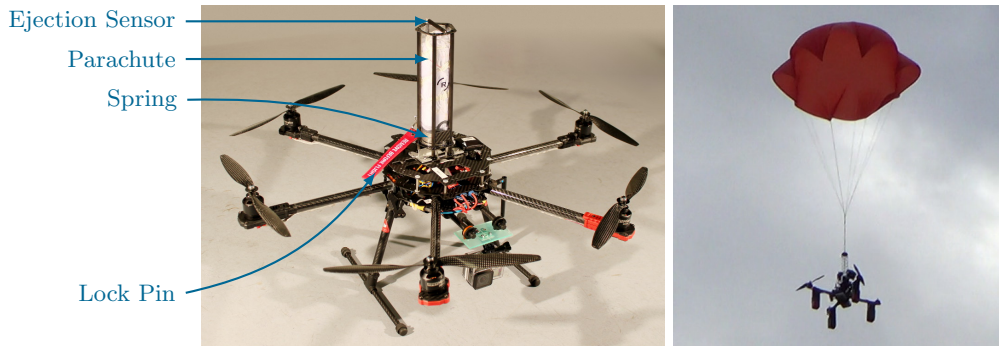

Whatever solution is chosen to increase the level of safety, it has to
be tailored towards those low-cost, mass-market systems. Imposing
certification rules on the entire MAV, such as DO-178C for civil A/C
software, could eventually hold back a number of desirable use
cases. For example, certification could require redundancy in the
flight controls, which would decrease payload capacity and thus render
some applications infeasible. Last but not least, low cost is also a
key for those platforms, which generally contradicts a full-system
certification.

In this paper, we describe our experiences in developing a
light-weight recovery system which increases the operational safety of
MAVs and is nevertheless amenable to certification, independently from
the internal structure of the MAV. It is a hardware-software solution
based on a parachute, which can bring down the MAV safely, avoiding
loss of the MAV in case of malfunctions, and minimizing collateral damage.
Our system is a ``plug and play'' solution, i.e., it can be
retrofitted to existing MAVs with only one single interface (the power
connector) and has little impact on the flight performance.

In the following we first explain the overall solution, and then focus
on the verification of the embedded software, which is the most
complex part, and meanwhile the main contributor towards the
effectiveness of the proposed solution.


\section{Related work}
\vspace{1mm}\textbf{MAV Safety Systems:} In general, the safety systems available
by today are either specific to the MAV brand, incomplete, or require
radical modifications to the existing MAV.  For example, there are
MAVs that ship with a parachute system, such as the
MCFLY-Helios~\cite{McFly}, or others which can be extended with OEM
parachute systems, such as the ``DropSafe'' for the \emph{DJI
  Phantom}~\cite{dropsafe14}. However, being tightly integrated with
their specific MAVs, the trigger conditions are not made public, and
there is no formal proof illustrating the increased overall
safety. Moreover, they require CO\textsubscript{2} capsules and a
backup battery, as opposed to our solution.
Other available systems are ``operated'' solutions, such as the
Opale~\cite{opale}, SKYCAT~\cite{skycat} or MARS~\cite{mars} parachute
systems. They only support a manual release, do not switch off the MAV
propulsion and require, as the others before, a working power supply
in case of emergency.

There are also more \emph{local} approaches to increase the safety of
subsystems, such as robust control algorithms by Mueller and D'Andrea
\cite{mueIEEE14}. Their algorithms can cope with partial loss of
propulsion whilst keeping the MAV in a controlled flight.  However,
not only do they require a lot of insight into and modification of the
MAV, they also demand significant non-local changes, as for example the
provisioning of safety margins in the propulsion (e.g., more thrust
per motor, higher peak current etc.).  Eventually, those margins make
the MAV inefficient under normal conditions, but still only
cover a subset of all possible MAV failures.

The parachute solution that we propose offers similar operational
limits than the mentioned automatic systems, but is MAV-independent
and covers the maximum number of failure conditions among the
mentioned solutions, and at a lower weight. Additionally, through the
verification shown here, we have evidence that the overall MAV safety
is indeed increased, as opposed to all other solutions.

\noindent\textbf{Verification of Code Running on Microprocessors:}
Model-checking the entire C code running on microprocessors has been
reported only a couple of times, e.g., with \emph{cbmc} on an ATmega16
processor in~\cite{Schlich2009} and on an MSP430 in~\cite{cbmcisr},
but either it failed because of state space explosion and missing support
for concurrency, or succeeded only for smaller programs.

However, recent developments that turn concurrency into data
non\-deter\-minism~\cite{Lal2009}, spot race
conditions~\cite{cbmc-racecond} and support for interrupts in
\emph{cbmc}~\cite{cbmcisr} can solve the concurrency issues and make
bounded model checking an interesting approach. In this paper we take
together all these ideas, point out problem with those, and propose
abstractions which mitigate the state space explosion, enabling a
workflow which allows verifying an entire real-world program running
on a microcontroller.








\section{Challenges}
The main design challenge for this system is to maintain a low weight, since
this directly translates into flight time. This however means we can
introduce redundancy only where inevitable for safety.  

Second, to make the system work independently of MAV internals, it
implies that the interface to the MAV must be minimalist. Standard
approaches known from avionics like \emph{triplex controllers} (see
\cite[p. 88]{Abzug05}) with its internal data consolidation are too
intrusive and therefore not an option.

The biggest challenge however, is deciding whether there is an
emergency, and triggering the recovery independently of the pilot. A
software implementation is the natural choice, since this allows for
iterative development and parametrization for the specific
MAV. This software is then \emph{safety-critical}, since it
directly influences whether crashes can be avoided or not. Through
this, the quality of the software will drive the quality of the
overall solution. That is why in this paper our main concern is a
formal verification of the software, which is known to be
challenging, especially because this software interacts with its
physical environment.



\section{Proposed Emergency Recovery System for MAVs}
Our proposed Emergency Recovery System (ERS) is shown in
Fig.~\ref{fig:deploy}, both on a quadrocopter and a hexacopter. It is
a parachute system, designed to increase the overall safety of
the MAV. In case of an emergency (what constitutes an emergency is
described later), the ERS automatically turns off the propulsion and
deploys a parachute. The technical specifications are given in
Table~\ref{tab:specs}.


\bgroup
\vspace{-1.25em}
\setlength\tabcolsep{.6em}
\begin{table}
\small
  \centering
  \caption{Specifications of the Emergency Recovery System.}
  \vspace*{-1em}
  \begin{tabular}{ll}
    \hline
    Property   & Value  \\
    \hline
    total weight   & \unit[320]{g}      \\
    input voltage    & \unit[6\dots25.2]{V} (2\dots6 LiPo cells) \\
    power consumption  & \unit[$<3$]{W} depending on propulsion state   \\
    worst-case trigger time  & \unit[$\leq$140]{ms}      \\
    terminal speed \& min. altitude  & \unitfrac[4.5]{m}{s} within \unit[10]{m} \\
    \hline
  \end{tabular}
  \label{tab:specs}
\end{table}
\vspace{-1.25em}
\egroup

No modifications to the existing MAV are required, e.g., neither
altering the flight controller nor the propulsion system. Our system
effectively acts as a power proxy between MAV battery and MAV. The
only (necessary) interface for our ERS is the power connector, which
is why we call it a ``plug and play'' solution. A second optional
interface is for one RC channel, allowing the pilot to trigger the
parachute manually.





\subsection{Internal Structure}
\begin{figure}[btp]
\vspace*{-5mm}
  \centering
  \includegraphics[width=.9\textwidth]{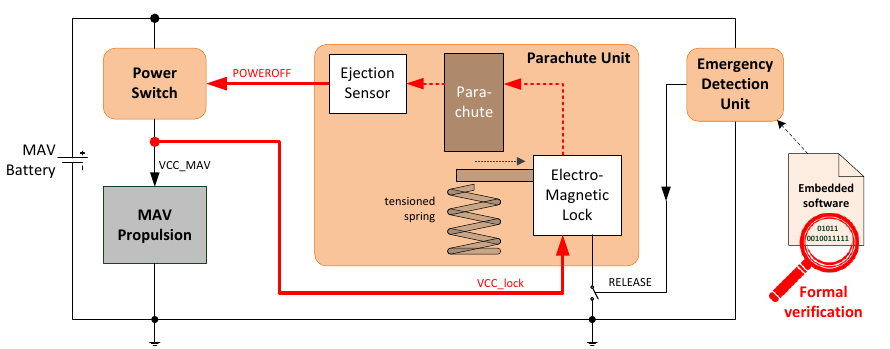}
  \vspace*{-1.5em}\caption{Internal Structure of our \emph{Emergency Recovery System}: The \emph{Emergency Detection Unit} on the top right is running the to-be-verified software.
    \label{fig:internals}}
\end{figure}

The ERS consists of the following three components, also illustrated
in Fig.~\ref{fig:internals}: \vspace*{-.25em}
\begin{itemize}
\item \textbf{Emergency Detection Unit (EDU):} A Printed-Circuit Board (PCB) with sensors and a
  microprocessor running software to detect emergencies. In
  case it detects an emergency, it can trigger the ejection of the
  parachute. 
\item \textbf{Power Switch (PS):} A PCB with power electronics, acting
  as a proxy between the MAV's battery and the propulsion. In case of
  emergency, it cuts off the power.
\item \textbf{Parachute Unit (PU):} This is a housing holding the
  parachute. It is also comprising an ejection sensor and an
  electro-magnetic (EM) lock, which, when opened or powerless,
  releases a compressed spring, which in turn ejects the parachute. \vspace*{-.25em}
\end{itemize}

\noindent\textbf{Mode of Operation:} 
The EDU features an Atmel ATmega 328p microprocessor (Harvard,
\unit[8]{MHz}, \unit[32]{kB} Flash, \unit[2]{kB} RAM, no caches), a
barometer sensor and an accelerometer sensor.  The embedded software
evaluates those sensors periodically, and estimates the MAV's air
state. When it detects emergency conditions, it triggers the parachute
ejection by emitting a \texttt{RELEASE} signal, which opens the EM
lock. This releases a compressed spring, which can now eject the
parachute from its housing.  Simultaneously, when the parachute is
pushed out, an ejection sensor detects this and sends a
\texttt{POWEROFF} signal to the \emph{Power Switch}. This ensures,
that the MAV's propulsion is deactivated as soon as the parachute is
ejected.

%
%
%
\noindent\textbf{Emergency Conditions:} The root causes
for failure in MAVs are wide-spread. Due to tight integration of
functionality and -- as explained before -- the imperative minimalism
in redundancy, even errors in non-critical components can evolve
quickly into fatal failures. Therefore, it seems more efficient to
apply a holistic monitoring, instead of monitoring single
components. Accordingly, an emergency is considered as the MAV being
\emph{uncontrolled}, that is, when the pitch or roll angles exceed
user-defined thresholds, or when the descent rate gets too
high.  
%
These conditions cover the most important malfunctions, such as FCS
failure (e.g., badly tuned controllers or error in software logic),
electrical or mechanical failure of propulsion (propeller, ESC), loss
of power and partially even human error (in the form of initiating an uncontrolled state).

\section{System Level Failure Analysis}
Although this paper focuses on software verification, we briefly
explain the failure analysis at system level, to show the influence of
the software on the overall safety.

We designed our ERS to make it fail-safe together with the MAV
w.r.t. any \emph{single-failure} event, i.e., a MAV equipped with our
ERS can tolerate at least one statistically independent failure
without leading to a crash. Towards that, we repeatedly conducted a
Fault Tree Analysis during the design process of the ERS.

\begin{figure}
  \centering
  \includegraphics[width=1.0\textwidth]{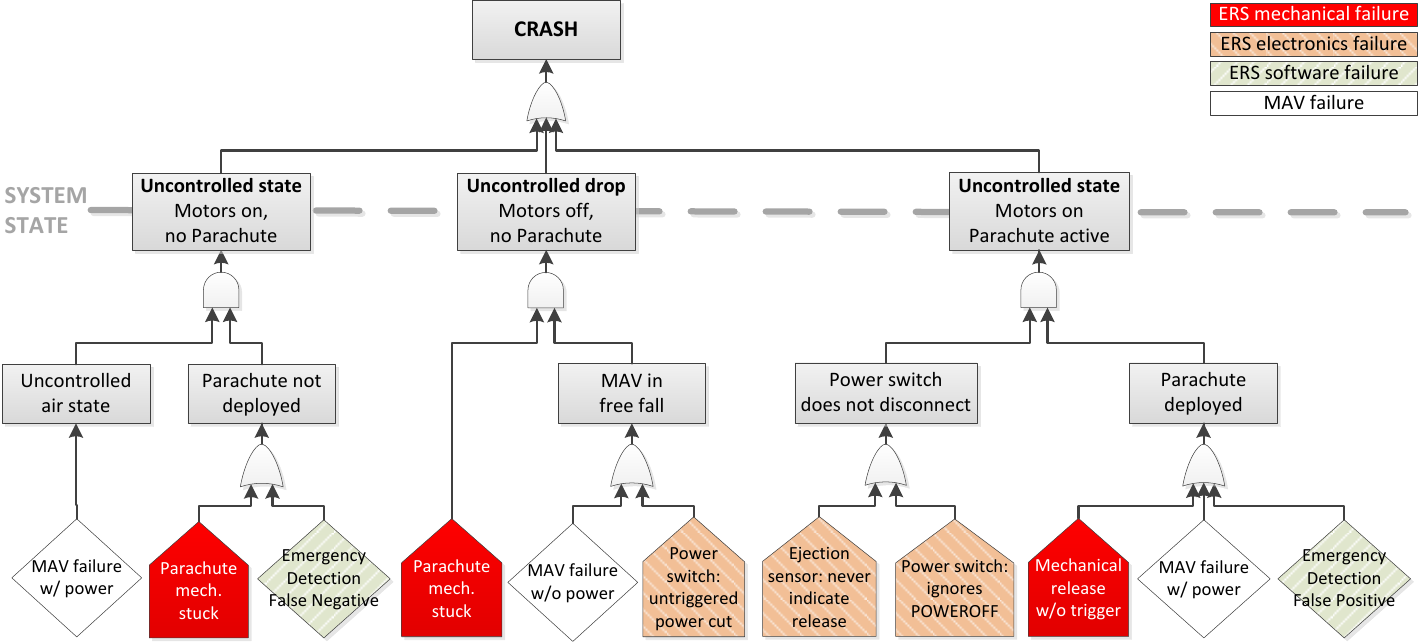}\vspace{-.5em}
  \caption{Fault Tree for the top event ``crash'', valid for any electric Micro Air Vehicle equipped with our Emergency Recovery System.\label{fig:fta}}
\end{figure}

In Fig.~\ref{fig:internals} we highlighted a built-in \emph{fail-safe
  loop} between power switch, EM lock, parachute and ejection
sensor. It creates a circular dependency between its components. If
any of them fails (e.g., broken power switch), then this also leads
to the ejection of the parachute, thus covering failures that may
occur in the ERS itself. The effects of different failure scenarios
can be seen in the Fault Tree in Fig.~\ref{fig:fta}.

\noindent\textbf{Considered MAV Failures}: The MAV was treated as a black box
with two possible failures (grey in the figure) ``MAV failure with
power'' and ``MAV failure without power''. The first one means, that
the MAV is in an uncontrolled state but still powered (e.g., broken
propeller and resulting loss of control), whereas the latter one
means, that the MAV lost power (e.g., due to battery failure or
electronic defects), which naturally results in an uncontrolled state
as well. We are not concerned with the MAV being powered up in a
controllable state (no error), or being in a controllable but
unpowered state (impossible for multicopter configurations).

\noindent\textbf{Influence of the Software:} The Fault Tree is depicted in
Fig.~\ref{fig:fta}. It can be seen, that the three \emph{uncontrolled}
system states which lead to a crash, can only be reached if at least two
failures occur at the same time. As indicated with the color coding,
there are four categories of failures:
\begin{inparaenum}[\itshape a\upshape)]
\item mechanical failure in ERS (red),
\item electronics failure in ERS (orange),
\item software failure in ERS (green) and
\item MAV failure (white).
\end{inparaenum}
%
Although there are many kinds of
errors possible in software, from a system point of view we are only
interested in the two consequences depicted in the Fault Tree:
\vspace*{-.5em}
\begin{enumerate}
\item \textbf{Emergency Detection False Negative}: The embedded
  software does not trigger the emergency sequence despite emergency
  conditions.
\item \textbf{Emergency Detection False Positive}: The embedded
  software does trigger the emergency sequence without emergency
  conditions.
\end{enumerate}

\noindent While both software failure events can have the same impact at system level
(both can lead to crash if a second failure occurs), the case of a
\emph{False Negative} is practically more critical, since MAV failures
with power are more likely than a second independent failure occurring
in the ERS. Furthermore, the ERS runs self-checks during
initialization, reducing the probability of being used in the presence of
internal failure. For these reasons, our verification efforts that we
explain in the next section, focused on (but were not limited to)
finding defects that lead to False Negatives.

\section{Software Verification}
Safety-critical systems in general must be free of defects that can
lead to errors in behavior. Here, traditional testing is not
favorable, since only a full coverage of all possible executions could
guarantee absence of defects, which implies modeling the system's
environment in a test harness.
That especially holds true for our ERS, where the functionality
strongly depends on timing and the interaction with its
environment. Testing specific cases would require simulating the
environment, as well as the sensors and the microprocessor running the
software. On top of that, in our system we cannot afford any
redundancy due to weight reasons, which is why we need to identify all
defects in the software.


Consequently, we aimed for a toolchain that supports formal
verification of C code based on static analysis. While there are multiple
tools that one could choose for that task (e.g., Frama-C
\cite{FramaC12}, Astr\'ee \cite{astree05}, BLAST \cite{blast03},
Polyspace, etc.), we have selected \emph{cbmc} and related tools
\cite{ckl2004}, because they support concurrency to some extent, are
freely available (and thus can be extended if necessary) and also
widely used.
More model checkers for C code were compared in
\cite{Schlich2009} and \cite{svcomp14}.

\vspace{.5em}
\noindent\textbf{Software Structure:} The software running on the EDU
can be partitioned into four sequential parts:
\begin{enumerate}
\item \textbf{Initialization:} Initializes all sensors, and captures
  environmental conditions (e.g., pressure at ground level). When
  completed, the ERS switches to \emph{self-check mode}.
\item \textbf{Self-Check:} To ensure that there is not already a
  failure in the ERS during start-up, we added built-in self tests
  covering the major subsystems of the ERS.
  When completed, the ERS switches to \emph{detection mode}.
\item \textbf{Detection:} The software periodically reads all sensors
  and estimates the MAV's air state. If the emergency conditions
  apply, the EM lock is released and the software switches to
  \emph{emergency handling mode}.
\item \textbf{Emergency Handling:} Current sensor data and decision
  conditions are written to EEPROM, to enable a post-flight analysis.
\end{enumerate}

\begin{figure}
  \centering
  \includegraphics[width=1.0\textwidth]{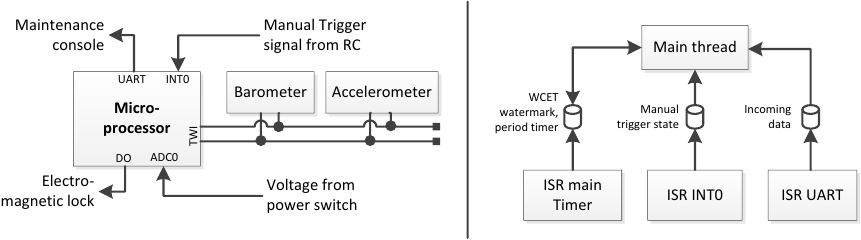}\vspace{-.5em}
  \caption{Microprocessor with interfaces to its environment (left) and the resulting concurrency in the software (right).\label{fig:threads}}\vspace{-.3em}
\end{figure}

The sensors and actuators are connected to the microcontroller as
depicted in Fig.~\ref{fig:threads} on the left. The interfaces impose
some concurrency in the software, which is shown on the right. For
example, the maintenance console and manual trigger signal both
require interrupts (polling would be too slow), thus each introduces
one thread concurrent to the main program. Additionally, a timer
interrupt is used to support a time-triggered execution of the
detection loop, contributing one further thread.


\noindent\textbf{Proper Timing:} The mentioned concurrency poses the first verification task. To ensure that the detection
loop always runs at the desired rate -- which is important for
correctness of computed data, e.g., the descent rate -- we need to
show that the required computations can be completed before the
next period begins.

Towards that, the \emph{worst-case execution time} (WCET) of the main
loop must be determined. Here we took a dual approach: On one
hand, we performed a static WCET analysis with a freely available analyzer
tool \cite{boundt}, but we also monitor the execution time on the microprocessor with a \emph{high watermark}. 

For the static analysis we made the assumption that the sensors are
healthy, and follow their datasheets' timing specification. The
resulting WCET was \unit[2.7]{ms} for the detection loop, which is
well below the \unit[5]{ms}-period in the EDU. However, interrupts
also need to be considered. The \emph{worst-case response time} (WCRT)
is (in this context) the maximum amount of time that the detection
loop needs to finish processing, under the preemption of
interrupts. Only if the WCRT is less than the period, then it can be
concluded that the timing is correct.

However, without further provisions the minimum inter-arrival time
(MINT) for the event-based interrupts (manual trigger from RC, UART)
have no lower bound, i.e., it would be possible that a broken RC
receiver or UART peer could induce so many interrupts, that the
detection could never execute, resulting in an unbounded WCRT. To
avoid this situation, the inter-arrival times of all event-driven
interrupts are also measured in the microcontroller. If an interrupt
occurs more often than planned, the attached signal source is
considered failing, and the interrupt turned off.

With these bounded MINTs and the WCET values from the static analysis,
a standard response time analysis yielded a WCRT of \unit[2.89]{ms}
for the detection loop. Again, this is for the case of healthy sensors.

The purpose of the high watermark is to detect those cases when
sensors are failing, but also to gain confidence in the above
analysis.  The response time of the detection loop is continuously
measured using a hardware timer, and maximum values are written to
EEPROM. With rising number of flying hours, the watermark should
approach the WCRT. If it exceeds the statically computed WCRT, then a
sensor failure is likely, which triggers the emergency sequence.

In practice, the watermark measurements were observed approaching the
statically computed WCRT up to a few hundred microseconds with healthy
sensors, thus giving confidence in the analysis. By construction of
the software, it can be concluded that the timing of the detection
loop is correct, unless the parachute is deployed. However, there are
more timing-related issues to be considered, namely, the time-sensitive effects of
interrupts upon the control flow in the main program. This was
addressed later during the verification process.

\noindent\textbf{Proper Logic:} The ultimate goal of the software
verification is to ensure that the emergency detection algorithm works
as intended.  As explained before, the main concern was to avoid False
Negatives, i.e., the error that the embedded software does not trigger
the emergency sequence, despite emergency conditions.

An obvious reason for such failure is, that the software is not
running because it crashed or got stuck. This can be a consequence of
divisions by zero, heap or stack overflow\footnote{Heap was not used,
  and stack size was checked with \emph{Bound-T}.}, invalid memory
writes, etc. Note that a reboot during flight is not possible, since
the initialization and self-checks need user interaction (open and
re-close the ejection sensor to ensure it works correctly), and making
them bypassable is not desirable for practical safety
reasons. Therefore, crashes and stuck software have to be avoided.

The second reason for not recognizing an emergency is an incorrectly
implemented detection algorithm. This entails both an error in
decision taking (i.e., which sensor has to tell what in order to
classify it as emergency), and also numerical problems (e.g.,
overflows) in data processing. Identifying these kinds of problems also
decreases the number of False Positives.

The majority of those defects is checked automatically by \emph{cbmc},
if requested during instrumentation. The correctness of the decision
taking part, however, must be encoded with user assertions. Since our
detection loop runs time-triggered, properties such as ``latest
\unit[100]{ms} after free fall conditions are recognized, the
parachute shall be deployed`` can be encoded with some
temporary variables. With that, verification of arbitrary properties
of the decision algorithm follows the same workflow as the
automatically instrumented properties, which is why we do not
elaborate on the specific properties that were eventually verified,
but rather show how we set up the workflow correctly.
 
\subsection{Verification Workflow}
\begin{figure}
  \centering
  \hspace*{-1.5mm}\includegraphics[width=1.025\textwidth]{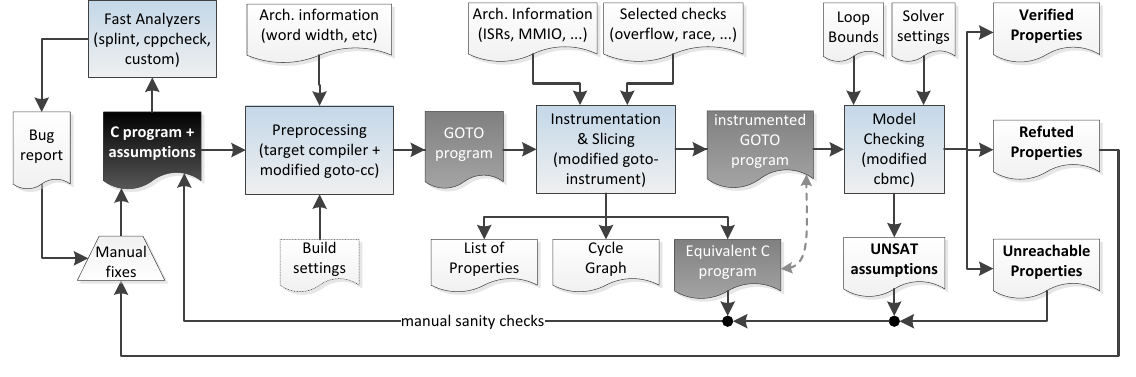}\vspace{-1em}
  \caption{Workflow for formal verification of the embedded software written in C.\label{fig:veri}}
\end{figure}

The toolchain that we set up around \emph{cbmc} is shown in
Fig.~\ref{fig:veri}. We start with a C program, written for the
target. First, we run fast static checkers such as \emph{splint} on
the program, to identify and remove problems like uninitialized
variables, problematic type casts etc. 
Not only does this help to avoid defects early during development and
thus to reduce the number of required verification runs later on, but
also it complements the verification. For example, the semantics of an
uninitialized variable depends on the compiler and the used operating
system (if any); \emph{cbmc}, however, regards these variables as
nondeterministic and therefore overapproximates the program without a
warning.

After passing the fast checks, the C code is given to \emph{goto-cc},
which translates it into a \emph{GOTO-program}, basically a control
flow graph. During this process, all the macros in the C code are
resolved by running the host compiler up to the preprocessing stage.

The \emph{GOTO-program} is subsequently fed into
\emph{goto-instrument}, which adds \emph{assert} statements according
to user wishes. For example, each arithmetic multiplication can be
checked for overflow, array bounds can be ensured, etc. Note that the
original code may contain user-defined assert statements, which are
preserved.

The resulting \emph{instrumented GOTO-program} is finally handed over
to \emph{cbmc}, which performs loop unwinding, picks up all
\emph{assert} statements, generates VCCs for them and -- after
optional simplifications such as slicing -- passes the problem to a
solver back-end (we use \emph{MiniSat2}; SMT solvers like \emph{Z3} and \emph{Yices}, are recent additions to \emph{cbmc}).

After the back-end returns the proofs, \emph{cbmc} post-processes them
and provides a list of verified properties, and for each refuted one a
counterexample. These lists can be used to fix defects in the original
code, clearing the way for the next iteration.

\subsection{Missing Architectural Information}
A problem in static verification is implicit semantics that depends on
the target, for example that certain functions are set up as interrupt
service routines (ISRs) and thus their effect needs to be considered,
although they never seem to be invoked. Another example is
memory-mapped I/O, which may seem like ordinary reads from memory, but
in fact could inject nondeterministic inputs from the environment.

Neglecting such context can easily lead to a collapsing verification
problem and result in wrong outcomes. In our program, there were
initially 351 properties, from which 349 were unreachable due to
missing contextual information. Annotating all the necessary places
manually is an error-prone labour, which bears the risk of having
wrong or missing annotations and more importantly it is practically
infeasible for our small program already. In the following we discuss
how we addressed this problem.

\noindent\textbf{Accounting for Interrupts:} The preprocessed C code
contains the ISR definitions, but naturally no functions call to
them. The ISR is only called because its identifier is known to the
cross compiler, and because particular bits are being written to
registers at the start of the program; something that the
model-checker lacks knowledge of.  Consequently, it concludes that the ISR
is never executed, and -- through data dependencies -- our detection
algorithm seems to be never executed. This makes all properties within
that algorithm unreachable and thus incorrectly evaluates them as
``verified''.

To overcome this, a nondeterministic invocation of the ISR must be
considered at all places where shared variables are being evaluated,
as described in \cite{cbmcisr}. This can be done with
\emph{goto-instrument} as a semantic transformation (flag
\texttt{--isr}). Fig.~\ref{fig:threads} shows the respective data that
depends on interrupts in our case.
Unfortunately, this technique not only grows the to-be-explored state
space, but it even overapproximates the interrupts: The ISR could be
considered too often in the case when the minimum inter-arrival time
is longer than the ``distance'' of the nondeterministic calls (e.g.,
ISR for periodic timer overflow) that have been inserted. However,
even if we would include execution time and scheduling information
from parts of the main thread (to be computed by WCET and WCRT tools),
the points in time where the ISR is called could be drifting w.r.t. to
the main thread. This is true even for perfectly periodically
triggered programs, solely due to different execution paths in the
main thread.

\noindent\textbf{Nondeterminism from Frequency-Dependent Side Effects:}
There exists another problem with interrupts that has not been
addressed in \cite{cbmcisr} nor in \texttt{goto-instrument}. It stems
from the frequency-dependent side effects of ISR invocation: In
general, interrupts could also execute \emph{more often} than the
places where nondeterministic calls have been considered before. If
there exist side effects other than changes to shared variables (i.e.,
if the ISR is non-reentrant in general), this can break the correct
outcome of the verification. For example, ISRs that on each invocation
increment some counter variable which is \emph{not} shared with any
other thread, could then in reality have a higher counter value than
seen by the model checker\footnote{A \emph{lower} value is not
  possible, because all considered invocations are nondeterministic
  possibilities, and not enforced invocations.}. In other words, all
persistent variables that are manipulated by the ISR have to be
modeled as nondeterministic, not only shared variables. In our case
there were only three such variables (one was for the time-triggered
release of the detection loop), which have been identified and
annotated manually.

\noindent\textbf{Memory-Mapped I/O:} All I/O variables (the sensor inputs) must
be annotated to be nondeterministic. One option for that would be
using the flag \texttt{--nondet-volatile} for \emph{goto-instrument}
to regard all volatiles as nondeterministic, however, this results in
overapproximation for \emph{all} shared variables (which are volatile
as well), allowing for valuations which are actually infeasible due to
the nature of the algorithms operating on the shared
variable. Furthermore, this can override user-defined assumptions on
the value domain of sensors, considering actually impossible
executions and thus produce False Negatives on the verified
properties.


In our case the microcontroller runs bare-metal code and uses
memory-mapped I/O to read sensors, i.e., accesses show up in the
preprocessed C code as dereferencing an address literal. In principle,
it is therefore possible to identify such reads after the C
preprocessing stage.
However, in general it is a non-trivial problem to identify all these
places, since indirect addressing is possible, which would require a
full value analysis of the program to figure out whether the effective
address is in the I/O range.  At the moment we do not have a practical
solution to this problem, which is why we instrumented all inputs
manually. To support this process, we developed a
\emph{clang}-based~\cite{lattner2008llvm} tool which generates a list
of all dereferencing operations, suggesting the places that should be
considered for annotating nondeterminism in the C code. Since we
minimized the use of pointers to keep verification effort lower, the
majority of the entries in this list is indeed reading input
registers.



\subsection{Preprocessing against State-Space Explosion}
After all architectural information has been added, the next big
challenge is to verify the instrumented properties. A problem here is,
that the state space grows rapidly from the architectural features,
especially from the ISRs. In our case, the program has around 2,500
lines of C code, and running \emph{cbmc} already fails for two
reasons: (1) the program contains unbounded loops and (2) even if the
loops were somehow bounded, there would be too many SAT variables to be
considered (millions in our case).


\noindent\textbf{Building Sequential Modes:} The original structure of our
program could not be verified, because the initialization and
self-checks, were implemented as part of one hierarchic state machine,
executed in main loop. The necessary loop unwinding then expanded the
entire state machine as a whole. This resulted in too many SAT
variables and could not be processed on our machine (we run out of
memory after hours, having done only a fraction of the necessary
unwinding).

\begin{figure}
\vspace*{-1.6em}%
  \centering
  \includegraphics[width=1.0\textwidth]{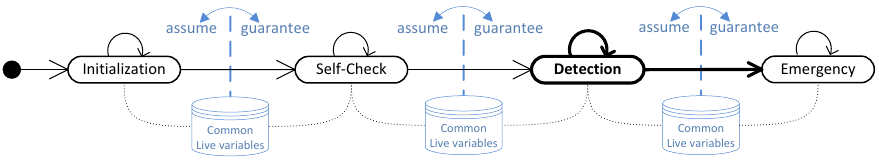}\vspace*{-.9em}
  \caption{Partitioning of software into strictly sequential modes,
    each verified individually and cascaded using
    \emph{assume-guarantee} reasoning.\label{fig:modes}}\vspace{-1.9em}
\end{figure}

To overcome this state space problem, we first partitioned our program
into sequential modes, see Fig.~\ref{fig:modes}. Each the
initialization, the self-tests and the detection were refactored into
their own loops, which take place one after another. Interrupts were
enabled as late as possible, reducing the number of states to
explore.

\noindent\textbf{Assume-Guarantee Reasoning}: However, at this point it turned
out, that the initialization and self-checks still contributed too many
variables for the program to be analyzed as a whole.  As a
countermeasure, the modes should now be analyzed independently and
reasoning on the overall correctness should be done using
\emph{assume-guarantee} reasoning. Towards that, it was necessary to
identify all possible program states between the modes, e.g.,
the detection mode can only be properly analyzed, if all possible program
states after initialization and self-check are considered. One
concrete example is, that the ERS determines the air pressure
at ground level during the initialization, which is used later during
detection. Verifying the detection mode thus involves considering all
possible pressure levels, by assuming nondeterministic values for
them.

To reduce the complexity of assume-guarantee
reasoning, we first turned each mode into a potentially infinite loop
which can only exit, if everything works as expected 
These ``guards'' reduce the number of program states to be considered for
the postdecessor modes. For example, when analyzing the detection
mode, we only need to consider program states corresponding to 
\emph{successful} initialization and self-checks.

To construct the program states between modes, we
identified all \emph{live variables} between each two successive
modes, i.e., all variables which are written in one mode and possibly
being read in its successor modes.
As this is another error-prone work that should not be done manually, we
extended our \emph{clang}-based tool to take this step automatically.

 After having identified the live
variables at the end of each mode, we instrumented them as illustrated
in Listing~\ref{lst:ag}: First, we added a nondeterministic
assignment to each variable just before the new mode starts (line
6). This allows for \emph{all} possible values, once the analysis on the new
mode starts. Then, if due to some logical reason the value range could
be limited, we used an \texttt{assume} statement to restrict analysis
to this value range (line 7).  However, to \emph{guarantee} that the
value domain is indeed complete, i.e., ensuring that no possible
execution has been neglected, we added a matching \texttt{assert}
statement at the exit of the predecessor mode (line 3).

\lstset{caption={Illustration of assume-guarantee reasoning using
    \emph{cbmc} at the program point between two sequential modes
    \texttt{X} $\rightarrow$ \texttt{Y}, sharing one live variable
    \texttt{sharedvar}.},label=lst:ag}
\begin{lstlisting}
// end of mode X
#ifdef ANALYZE_MODE_X
  assert(sharedvar > -10.f && sharedvar < 50.f);
#endif
#ifdef ANALYZE_MODE_Y
  sharedvar = nondet_float(); // introducing nondeterminism
  assume(sharedvar > -10.f && sharedvar < 50.f);
#endif
// beginning of mode Y
\end{lstlisting}

A successful verification of the predecessor mode (here: \texttt{X})
means the \texttt{assert}s hold true, therefore \emph{guarantees} that
live variables indeed satisfy the assumptions we make at the beginning
of the new mode (here: \texttt{Y}). Assume-guarantee reasoning
therefore is sound. Finding the value ranges is currently done
manually; in doubt one can omit the ranges, which leads to a safe
over-approximation. However, tool support would be favorable, since
tight ranges means no false alerts during verification.





In summary, this mode-building reduced the number of properties from
458 to below 250 in each mode, with 31 shared variables between them
that were subject to assume-guarantee process (see Table
\ref{tab:complexity}).

\noindent\textbf{Removing Dead Code:} 
When going through the verification process shown in
Fig.~\ref{fig:veri}, it is desirable to entirely remove dead code
(especially after mode-building and analyzing the modes separately),
otherwise a lot of unreachable properties will be there, slowing down
the analysis and cluttering the results. Although
\emph{goto-instrument} offers two slicing options, none of them
removes dead code. This task is not trivial, since in our case the
modes share code, e.g, both self-check and detection use a function
that reads out the accelerometer. Again, we used our clang-based tool
for this task, which operates on the C code that is equivalent to the
GOTO-program and removes dead functions and variables (see Fig.~\ref{fig:veri}). 



\noindent\textbf{Bounding Non-local Loops}: A
complexity-increasing problem for verification are nested, stateful
function calls, as they occur in hierarchical state machines. Our
program uses such hierarchical state machines to interact with the
barometer and accelerometer peripherals. If one of the inner states
has transition guards, then the \emph{entire} hierarchy needs
unrolling until these guards evaluate to true. In our case, we have
guards like \emph{waiting for ADC conversion to finish}.
Unfortunately, hierarchic state machines are a popular design pattern
in model-based design (e.g., Statemate, Stateflow, SCADE), which
therefore needs to be addressed rather than avoided.

We found that some guards in the inner state machines can be removed
safely, reducing costly unrolling. Assume that the guard will
eventually evaluate to true (even if there is no upper bound on the
number of steps it takes): If all \emph{live} data that is written
\emph{after} this point is invariant to the number of iterations, then
the guard can be removed. Consequently, such irrelevant guards can be
identified by first performing an impact analysis (find all variables
that are influenced by the guard), followed by a loop invariance test
(identify those which are modified on re-iteration) followed by a live
variable analysis on the result (from the influenced ones, identify
those which are being read later during execution). If the resulting
set of variables is empty, then the guard can be removed safely. This
technique is of great help for interacting with peripherals, where
timing may not influence the valuations, but otherwise contribute to
state space explosion.  The technique is easily extended, if there are
multiple guards.

On the other hand, if a guard potentially never evaluates to true,
e.g., due to a broken sensor, then there are two ways to treat this:
If this is valid behavior, then this guard can be ignored for the
analysis (no execution exists after it). If it is invalid behavior,
then the guard should be extended by an upper re-try bound and this
new bounded guard can then be treated as explained above. After these
transformations all state machines could be successfully unrolled.

\bgroup
\setlength\tabcolsep{.6em}
\begin{table}[btp]
\begin{minipage}{\linewidth}
  \centering
  \caption{Complexity of the verification before and after
    preprocessing. Unlike the full program, which cannot be analyzed,
    assume-guarantee reasoning between sequential modes
    \emph{Initialization}, \emph{Self-Check} and \emph{Detection} was
    computationally feasible.  \vspace*{-2mm}}
    \small
  \begin{tabular}{lrrr|r}
    \hline 
    Mode $\rightarrow$   & {Initialization} & {Self-Check} & {Detection}  & {All}\\
    \hline
    lines of code            & 1,097         & 976           & 1,044   & 2,513 \\
    \#functions              & 36            & 29            & 43      & 94\\
    \#persistent variables   & 36            & 38            & 59      & 72\\
    \#live variables at exit & 31            & 31            & n.a.    & n.a. \\ 
    \hline
    \#properties             & 249           & 221           & 175       & 458 \\
    \#VCCs                   & 11,895        & 35,001        & 15,166    & 330,394 \\    
    \#SAT variables          & 5,025,141     & 8,616,178     & 6,114,116 & n.a. \\    
    SAT solver run-time\footnote{On an Intel Core-i7 vPro at \unit[2.8]{Ghz} and \unit[4]{GB} RAM.} & \unit[16]{min} & \unit[14]{min}  & \unit[28]{min} & infeasible\footnote{Out of memory after 3 hours; \#VCCs and SAT variables were still growing.} \\
    \hline
  \end{tabular}\label{tab:complexity}
\end{minipage}\vspace*{-.5em}
\end{table}
\egroup

\subsection{Keeping Assumptions Sound}
We made use of assumptions for limiting value domains where possible,
and to perform assume-guarantee reasoning. Assumptions are a powerful
tool in \emph{cbmc}, however, it is easy to add assumptions which
are not satisfiable (UNSAT). Those rule out \emph{all} executions
after the \texttt{assume} statement and thus might lead to wrong verification
results.

Therefore, we have to ensure that the \emph{composite} of all
annotations is sound, otherwise the verification outcome may be wrong
despite the individual annotations being correct.
To check whether assumptions can be satisfied, we added a new check to
\emph{cbmc}, which does the following: It inserts an
\texttt{assert(false)} after each assumption and subsequently runs the
SAT solver on it. If the solver yields UNSAT for the assertion, it
means it is reachable and thus the assumption is valid. If it yields
SAT, then all executions were ruled out and thus the assumption is
UNSAT and thus unsound. Finally, we warn the user for each
UNSAT assumption.

\subsection{Verification Results}
With our extensions of existing tools we were able to set up a correct
verification workflow for the software of the ERS. The complexity of
the analysis (for each mode: run-time, number of variables etc.) is
summarized in Table \ref{tab:complexity}.  During the process we
identified several trivial and non-trivial defects, some of them were
one deadlock in a state machine, multiple overflows in sensor data
processing and even one timing-related error (barometer update took
more steps than anticipated, which lead to wrong descent rate).
Interestingly enough, during flight tests we sporadically experienced
some of these errors, which by then could not be explained. One of the
reasons for this is, that there was little information about these errors
due to limited logging and debugging facilities on the
microcontroller, and that we could not reproduce the environmental
conditions in the lab.

\section{Conclusion}
In this paper we described our approaches in developing a
safety-critical emergency recovery system for MAVs, in particular our
efforts in applying methods and tools for formal verification of
embedded software.
%
This study has shown that formal verification of the entire, original
software running on a microcontroller is possible, if appropriate
preprocessing techniques are applied. The state space can be reduced
to a size that can be covered by existing tools, but careful handling
is necessary to obtain correct results.  The efforts did pay off in
our case. Not only could we identify defects in the software, but we
obtained counterexamples for the defects, which can be the only useful
source of debugging information for resource-constrained embedded
systems.


As future work, we are planning to extend our clang-based tool to
perform not only some, but all the steps we have taken automatically,
as well as a complementary software supporting the described iterative
workflow.

{
 \bibliographystyle{splncs03}
 \bibliography{literature}
}

\end{document}